\documentclass[dvips,12pt]{article}

\usepackage{palatino,amsmath}

\newcommand {\half} {{1 \over 2}}
\newcommand {\hhalf} {{\textstyle{1 \over 2}}}
\newcommand {\hhhalf} {{\scriptstyle{1 \over 2}}}

\newcommand {\ket}[1] {\left| #1 \right>}
\newcommand {\braket}[2] {\left< { \left. #1 \,\right|} \,#2 \right>}
\newcommand {\Tr} {\mathrm{Tr}\,}
\newcommand {\im} {\mathrm{im}\,}
\newcommand {\sig} {\mathrm{sig}\,}
\newcommand {\Jpmarrow} {\genfrac{}{}{0pt}{1}{\xrightarrow{\phantom{n}J_+\phantom{n}}}
                         {\xleftarrow[\phantom{n}J_-\phantom{n}]{}}}

\begin{document}

    \title{{\footnotesize{\hfill BROWN-HET-1312}}
            \\On the Representation Theory of Negative Spin}
    \author{Andr\'e van Tonder
            \thanks{Work supported in part by DOE grant number DE FG02-91ER40688-Task A}
            \\ \\
            Department of Physics, Brown University \\
            Box 1843, Providence, RI 02906 \\
            andre@het.brown.edu}
    \date{July 11, 2002}

    \maketitle

    \begin{abstract}
        We construct a class of negative spin irreducible representations
        of the $su(2)$ Lie algebra.  These representations are
        infinite-dimensional and have an indefinite inner product.
        We analyze the decomposition of arbitrary
        products of positive and negative representations with the help of
        generalized characters and write down explicit
        reduction formulae for the products.
        From the characters, we define effective dimensions for
        the negative spin representations, find that they are
        fractional, and point out that the dimensions behave
        consistently under multiplication and decomposition of
        representations.
    \end{abstract}

    \section{Introduction}

    The representation theory of $su(2)$ is
    familiar to most physicists and mathematicians \cite{ELLIOTT}.  The
    finite dimensional representations are labelled by
    their spin, which is non-negative and can take either integer or
    half-integer values.

    Recently, we found that a certain infinite-dimensional
    representation of $su(2)$ of negative spin,
    with value $-\half$, appears in the description of
    a ghost particle
    complementary to a spin $\half$ degree of freedom
    \cite{MYSELF}.
    Although this Lie algebra
    representation did not exponentiate to give a
    representation of the whole $SU(2)$ group, we found it
    interesting to investigate in
    its own right.

    We found that
    the product of the spin $\half$ and spin $-\half$
    representations was
    just the spin $0$ representation, modulo a
    well-defined reduction procedure in the product space.  We
    showed that this cancellation of spins could be reflected in
    character formulae and argued that the spin $-\half$
    representation effectively had a fractional dimension equal to
    $\half$.

    In this paper, we extend these results to
    representations of arbitrary negative spin.  These
    representations are all defined on infinite-dimensional
    indefinite inner product spaces.

    We analyze the decomposition
    of arbitrary products of positive and/or negative representations.
    In general, we will see that the product representation
    is well-behaved only after carrying out a well-defined reduction procedure
    to the cohomology of a nilpotent operator $Q$ constructed in
    terms of
    the Casimir $\mathbf{J}^2$.

    We then define characters for the negative spin
    representations and, with the help of these, we write down
    explicit formulae for the reduction of general product
    representations.

    From the characters, we define effective dimensions for
    the negative spin representations, find that they are
    fractional, and point out that the dimensions behave
    consistently under multiplication and decomposition of
    representations.

    \section{The Lie algebra $su(2)$}

    In this section we summarize some basic formulas regarding $su(2)$
    that will be
    useful in what follows.

    The commutation relations of $su(2)$ are \cite{ELLIOTT}
    \[
        [J_x, J_y] = iJ_z,\quad [J_y, J_z] = iJ_x,\quad [J_z, J_x] =
        iJ_y.
    \]
    We can define ladder operators
    \begin{eqnarray}
        J_+ &\equiv &J_x + iJ_y, \nonumber \\
        J_- &\equiv &J_x - iJ_y  \label{LADDER}
    \end{eqnarray}
    satisfying
    \[
        [J_+, J_-] = 2 J_z.
    \]
    The Casimir operator
    \begin{eqnarray}
        \mathbf{J}^2 &=& J_x^2 + J_y^2 + J_z^2 \nonumber \\
        &=& J_+ J_- + {J_z}^2 - J_z \nonumber \\
        &=& J_- J_+ + {J_z}^2 + J_z
        \label{CASIMIR}
    \end{eqnarray}
    commutes with $J_x$, $J_y$ and $J_z$ and serves to label
    irreducible representations by their spin $j$ defined as $\mathbf{J}^2 =
    j(j+1)$.

    The lowest weight state, annihilated by $J_-$, of a spin $j$ representation
    has $J_z$ eigenvalue $m = -j$ and we denote it by $\phi_{-j}^{(j)}$.
    Labelling basis elements by
    the spin $j$ and
    $J_z$ eigenvalue $m$, the full set of basis vectors is generated by
    applying the ladder operator $J_+$ repeatedly to the lowest
    weight state to give the sequence
    \[
        \phi_{-j + k}^{(j)} \equiv (J_+)^k \,\phi_{-j}^{(j)}.
    \]

    \section{Negative spin}

    Let us consider the possibility of constructing hermitian representations
    of the $su(2)$ Lie algebra that have negative spin $j$, where $j$ is either
    half-integral or integral.

    As above, we start from a lowest weight state
    $\phi_{-j}^{(j)}$ annihilated by $J_-$
    and consider the ladder of states
    \begin{equation}
        \phi_{-j}^{(j)} \stackrel{J_+}{\longrightarrow} \phi_{-j+1}^{(j)}
        \stackrel{J_+}{\longrightarrow}
        \phi_{-j+2}^{(j)} \stackrel{J_+}{\longrightarrow} \cdots \label{LADDERSTATES}
    \end{equation}
    obtained by repeatedly applying $J_+$.

    First, we show that negative
    spin representations require an indefinite inner product space.
    In a hermitian representation, the ladder
    operator $J_-$ is conjugate to $J_+$, so that we can
    write the inner product of the state $\phi_{-j + k}^{(j)}$ with
    itself as follows
    \[
      \braket{\phi_{-j + k}^{(j)}} {\phi_{-j + k}^{(j)}} =
        \braket {(J_+)^k \phi_{-j}^{(j)}}
        {(J_+)^k \phi_{-j}^{(j)}}
        = \braket{\phi_{-j}^{(j)}}
        {(J_-)^k(J_+)^k \phi_{-j}^{(j)}}.
    \]
    Commuting the $J_-$ operators to the right, this becomes
    \begin{equation}
        \braket{\phi_{-j + k}^{(j)}} {\phi_{-j + k}^{(j)}}
        = k! \, (2j)\,(2j - 1)\cdots(2j-k+1)\,
        \braket{\phi_{-j}^{(j)}}
        {\phi_{-j}^{(j)}}.
    \end{equation}
    Since $j < 0$, we see that squared norm of
    the sequence of states in (\ref{LADDERSTATES})
    have alternating signs.
    In other words, the inner product is indefinite.

    We now choose the squared norm of the negative spin
    $j$ lowest weight state $\phi_{-j}^{(j)}$ to be equal to $1$.  Defining
    the normalized state
    \begin{equation}
        e_{-j+k}^{(j)} \equiv {1\over i^k\sqrt {k! \, |2j|\,|2j -
        1|\cdots|2j-k+1|}}
                \,\phi_{-j + k}^{(j)}
    \end{equation}
    so that
    \begin{equation}
        \braket{e_{-j+k}^{(j)}} {e_{-j+m}^{(j)}} = \delta_{km}\, (-)^{k},
        \label{INNERPROD}
    \end{equation}
    it is straightforward to check, using the commutation relations and
    taking careful account of signs and
    factors of $i$, that
    \begin{eqnarray}
        J_+  \,e_m^{(j)} &= &i\,\left\{ \left(j + m + 1\right)\,\left|j -
                m\right|\right\}^{\half}\,e_{m+1}^{(j)}
                \qquad m = -j, -j+1, \dots\nonumber\\
                &= &\phantom{i}\left\{ (j + m + 1)\,(j -
                m)\right\}^{\half}\,e_{m+1}^{(j)},\qquad (-1)^{\half} \equiv
                +i
                \label{JPLUSACTION}
    \end{eqnarray}
    and
    \begin{eqnarray}
        J_- \,e_{m+1}^{(j)} &= &i\,\left\{ (j + m + 1)\,|j -
                m|\right\}^{\half}\,e_{m}^{(j)}
                \qquad m = -j, -j+1, \dots\nonumber\\
                &= &\phantom{i}\left\{ (j + m + 1)\,(j -
                m)\right\}^{\half}\,e_{m}^{(j)},\qquad (-1)^{\half} \equiv
                +i.
                \label{JMINUSACTION}
    \end{eqnarray}
    These formulae are the continuation to negative $j$ of the
    corresponding formulae for positive spin representations.

    Notice that when $j$ is negative, the
    sequence of states $e_{-j + k}^{(j)}$ does not terminate.  These
    representations are therefore infinite dimensional.  However,
    we shall see later that one can assign finite effective
    dimensions
    to the negative spin representations in a meaningful way.

    A peculiar feature of these negative spin representations is their asymmetry with
    respect to interchange of $J_-$ and $J_+$.  In particular,
    there is no highest weight state annihilated by
    $J_+$.

    With the indefinite inner product
    (\ref{INNERPROD}), the state spaces are examples of
    what is known in the literature as Kre\u{\i}n spaces (see
    \cite{BOGNAR}, \cite{MALCEV}, \cite{JACOBZYK} and references therein).

    \section{The spin $-\half$ ghost}

    In \cite{MYSELF} we constructed an $su(2)$ representation of
    spin equal to $-\half$ on the state space of a bosonic ghost
    complementary to a single two-state spinor degree of freedom.
    In this section we briefly review this construction.

    The ghost creation and annihilation
    operators $a^\dagger$ and $a$ satisfy
    \[
        [a, a^\dagger] = -1.
    \]
    The negative sign on the right hand side gives rise to an
    indefinite inner product on
    the Fock space generated by applying creation operators
    to the ground state $\ket{\half}_g$
    annihilated by $a$.

    A set of hermitian $su(2)$ generators
    is given by
    \begin{eqnarray}
        J_x &\equiv &\phantom{-}{\textstyle {i \over 2}} \left( \sqrt{N+1}\, a -
        a^\dagger
            \sqrt{N+1}\right),
        \nonumber\\
        J_y &\equiv &- {\hhalf} \left( \sqrt{N+1}\, a + a^\dagger
            \sqrt{N+1}\right),
         \nonumber \\
        J_z &\equiv & N + {\hhalf}, \label{LIE}
    \end{eqnarray}
    where $N \equiv -a^\dagger a$.
    It is easy to check that these generators satisfy the $su(2)$ algebra
    \[
        [J_x, J_y] = iJ_z,\quad [J_y, J_z] = iJ_x,\quad [J_z, J_x] =
        iJ_y.
    \]
    The ladder operators are
    \begin{eqnarray}
        J_+ &\equiv &J_x + iJ_y = -i\,a^\dagger \, \sqrt {N+1}, \nonumber \\
        J_- &\equiv &J_x - iJ_y = \phantom{-} i\,\sqrt {N+1}\,a, \nonumber
    \end{eqnarray}
    and satisfy $J_\pm^\dagger = J_\mp$ and
    \[
        [J_+, J_-] = 2 J_z.
    \]
    It is straightforward to calculate
    \[
        \mathbf{J}^2 = {\hhalf} \left( J_+ J_- + J_- J_+ \right) +
        {J_z}^2 = - {\textstyle {1 \over 4}}.
    \]
    Since $\mathbf{J}^2 = j(j+1)$, where $j$ denotes the spin, we
    see that
    \[
        j = -{\hhalf}.
    \]
    In other words, the representation that we have constructed
    has negative spin.

    Note that, since the entire state space is generated by
    applying the raising operator $J_+$ to the vacuum state, the
    representation is irreducible.  Also, in contrast to the positive
    spin irreducible representations, it is infinite-dimensional.

    The lowest weight state is the ground state $e^{(\half)}_{\half} = \ket{\half}_g$
    which has
    $J_z$ eigenvalue $m = 1/2 = -j$.
    There is no highest weight state.

    Taking the following phase convention for the basis
    elements
    \begin{equation}
      e^{(-\half)}_{m} = \ket{m}_g \equiv (-)^{m-\half}{1\over \sqrt {\left({m-\hhhalf}\right)!}}\,(a^\dagger)^{m-\hhhalf}
      \ket{\hhalf}_g, \qquad m = \hhalf, {\textstyle {3\over 2}},
      {\textstyle {5\over 2}}, \ldots \label {BASIS}
    \end{equation}
    it is straightforward to check explicitly that
    the relations (\ref{JPLUSACTION}) and (\ref{JMINUSACTION}) are
    satisfied by this representation.  Here we have introduced the
    shorthand $\ket{m}_g \equiv e^{(-\half)}_{m}$ for the spin
    $-\half$ basis vectors.

    \section{Representation of finite $SU(2)$ elements}
    \label{FINITE}

    Since the negative spin Lie algebra representations have no
    highest weight state, we do not expect to be able to
    exponentiate these representations to obtain a representation
    of the full $SU(2)$ group.  In particular, it is easy to see that
    rotations by $\pi$ that take the unit vector $\mathbf{z}$ to $-\mathbf{z}$
    are not representable on the negative spin state space, since
    such rotations would normally exchange lowest and highest weight states.

    The conclusion is that the Lie algebra representation of the
    previous sections
    does not exponentiate to a give a representation of the full
    group.

    However, it is still possible to represent a restricted class
    of $SU(2)$ group elements.  For the spin $-\half$ case, this
    was done in reference \cite{MYSELF}.

    \section{Multiplication of positive spin representations}

    Given two positive spin irreducible representations of spins $j_1$ and
    $j_2$, their tensor product is itself a representation which
    can be written as a direct sum of irreducible representations
    as follows \cite{ELLIOTT}
    \begin{equation}
        R^{(j_1)}\otimes R^{(j_2)}
          = \sum_{J = |j_1 - j_2|}^{j_1 + j_2} R^{(J)}, \qquad
          j_1, j_2 \in \{0, \hhalf, 1, {\textstyle {3\over 2}}, 2,
       \ldots\}. \label{PRODPOSITIVE}
    \end{equation}
    The characters of the positive spin representations of $SU(2)$
    are given by
    \begin{equation}
        \chi^{(j)}(\theta) = \Tr e^{i\theta J_z} =
        e^{-ij\theta}(1 + e^{i\theta} + \cdots + e^{2ij\theta}) = {\sin \left(j + \half\right)\,\theta \over \sin
        \half\theta}, \label{CHARPOSITIVE}
    \end{equation}
    and satisfy the orthogonality relation
    \begin{equation}
        {1\over 2\pi} \,\int_0^{2\pi}
        d\theta\,\chi^{j_1}(\theta)\,
        \chi^{j_2}(\theta) \, (1 - \cos \theta) = \delta_{j_1 j_2}.
        \label{ORTHOGONALITY}
    \end{equation}
    We remind the reader that a character is a function on the conjugacy
    classes of a group.  For $SU(2)$,
    all rotations through the same angle $\theta$ are in the same class,
    irrespective of the direction of their axes.  The direct sum
    decomposition is reflected in the following algebraic
    relationship satisfied by the corresponding characters
    \begin{equation}
        \chi^{{j_1}}\, \chi^{{j_2}} =
           \sum_{J = |j_1 - j_2|}^{j_1 + j_2} \chi^{(J)}, \qquad
          j_1, j_2 \in \{0, \hhalf, 1,{\textstyle {3\over 2}},2,
             \ldots\}. \label{PRODCHARPOSITIVE}
    \end{equation}
    which, given that the dimensions of the representations are
    $D^{(j)} = \chi^{(j)}(\mathbf{1})$, is consistent with
    \begin{equation}
        D^{(j_1)\otimes(j_2)} = D^{(j_1)}\, D^{(j_2)} = \sum_{J = |j_1 - j_2|}^{j_1 + j_2}
        D^{(J)}.
    \end{equation}
    In the following, we will show that the analysis of the direct sum decomposition
    in terms of characters can be generalized to include the negative spin
    representations.

    \section{Combining spin $\half$ and spin $-\half$ representations}
    \label{HALFHALF}

    As a warmup, we study the product
    of the spin $\half$ and spin $-\half$ representations \cite{MYSELF}.
    We will discuss the
    sense in which the positive and negative spin representations cancel
    to give an effectively trivial representation.

    First, notice that there is a lowest weight state in the
    product state space given by
    \[
        \ket{0}\equiv\ket{\textstyle
        -\half}_{s}\otimes\ket{\hhalf}_{g},
    \]
    where we have introduced the shorthand $\ket{\textstyle{\pm \half}}_s \equiv
    e^{(\half)}_{\pm \half}$ for the spin $\half$ basis vectors.
    By definition, this state is annihilated by
    $J_- \equiv J_-^{(\half)} + J_-^{(-\half)}$.
    Using the relation
    \begin{eqnarray}
        \mathbf{J}^2 &=& J_+ J_- + J_z^2 - J_z\nonumber
    \end{eqnarray}
    it trivially follows that
    \[
        \mathbf{J}^2 \, \ket{0} = 0.
    \]
    We are therefore dealing with an $su(2)$ representation of
    spin $0$ in the product space.  What is unusual about this
    representation, however, is that $J_+ \ket{0} \ne
    0$.  However, this state has zero norm.
    In fact, this is true for all higher states in the ladder since
    \[
        J_+^{n} \ket{0} = i^n \,\sqrt {n}\,(n-1)! \, \ket{n},
    \]
    where the states
    \begin{eqnarray}
        \ket{n} &\equiv& \sqrt{n}\,\left(i\,\ket{-\hhalf}_s\otimes\ket{n+\hhalf}_g
                    +
                    \ket{\hhalf}_s\otimes\ket{n-\hhalf}_g\right),
                    \nonumber\\
    \end{eqnarray}
    are null.

    In general, we would like to truncate the ladder of states
    generated by $J_+$ as soon as we reach a null state.  The
    proper way of doing this is by noticing that the Casimir
    operator $\mathbf{J}^2$ is nilpotent.  We can therefore
    regard it as a BRST operator (see \cite{BRST} and \cite{HENNEAUX})
    and calculate its cohomology,
    which defines a natural reduction procedure on the state
    space. For an overview of the elements of BRST technology that
    we will use, see appendix \ref{BRST}.

    To see that $\mathbf{J}^2$ is nilpotent,
    note that in terms of the basis consisting of $\ket{n}$ and the
    additional null states
    \begin{eqnarray}
        \ket{\tilde n} &\equiv& {1\over
        2\sqrt{n}}\left(-i\,\ket{-\hhalf}_s\otimes\ket{n+\hhalf}_g
                    +
                    \ket{\hhalf}_s\otimes\ket{n-\hhalf}_g\right),
                    \label{NTILDEN}
    \end{eqnarray}
    it is not hard to
    calculate
    \[
        \mathbf{J}^2 \ket{\tilde n} = \ket{n}, \qquad \mathbf{J}^2 \ket{n} = 0,
    \]
    leading to the matrix representation
    \begin{equation}
           \mathbf{J}^2 = \left(
                \begin{array}{cccc}
                        0 &  &  &  \\
                          & \fbox{\ensuremath{\begin{array}{rr}
                                    0 & 1 \\
                                    0 & 0
                                  \end{array}}
                            }
                             &  & \\
                          &  &
                            \fbox{\ensuremath{\begin{array}{rr}
                                    0 & 1  \\
                                    0 & 0
                                  \end{array}}
                            }
                                &  \\
                          &  &  &  \ddots
                \end{array}
            \right). \label{Q2}
    \end{equation}
    In this basis, the inner product is represented by the matrix
    \[
        G \equiv \left(
                \begin{array}{cccc}
                        1 &  &  &  \\
                          & \fbox{\ensuremath{\begin{array}{rr}
                                      & 1  \\
                                    1 &
                                  \end{array}}
                            }
                             &  & \\
                          &  &
                            \fbox{\ensuremath{\begin{array}{rr}
                                      & -1  \\
                                    -1 &
                                  \end{array}}
                            }
                                &  \\
                          &  &  &  \ddots
                \end{array}
            \right)
    \]
    It is easily checked that $(\mathbf{J}^2)^2 = 0$ and that $\mathbf{J}^2$
    is hermitian
    with respect to the inner product $G$, as follows from
    \[
        \mathbf{J}^2 = (\mathbf{J}^2)^\dagger = G (\mathbf{J}^2)^+ G,
    \]
    where the dagger denotes hermitian conjugation with respect to
    the indefinite inner product $\braket{\cdot}{\cdot}$ on the state space and
    the $+$ sign denotes the usual matrix
    adjoint.

    We now specialize the BRST analysis to the case at hand,
    taking $Q = \mathbf{J}^2$.
    Since the generators
    $J_i$ all commute with $Q$, they are physical
    operators in the sense of appendix \ref{BRST}.

    We can therefore consistently reduce these operators to the
    cohomology $\ker Q/\im Q$ of $Q$
    and use (\ref{AREDUCED})
    to define the induced
    operators  $[J_+]$, $[J_-]$ and $[J_z]$ on the quotient space
    by
    \[
      [J_i] \,[\ket{\phi}] = [J_i \ket{\phi}], \qquad \ket{\phi} \in \ker Q.
    \]
    The cohomology consists of the single class $[\ket{0}]$,
    and is a one-dimensional, positive definite Hilbert space.
    The induced generators $\{[J_+], [J_-], [J_z]\}$ on this space
    are zero, corresponding to the trivial
    representation of $su(2)$.

    \section{Operators on indefinite inner product spaces}

    In order to address the product decomposition in the general case,
    we need to understand the behavior of the Casimir operator $\mathbf{J}^2$
    a little better.  With this in mind,
    we review a few
    general facts regarding hermitian operators on indefinite inner
    product spaces, also known as pseudo-hermitian operators (see \cite{BOGNAR},
    \cite{MALCEV}, \cite{JACOBZYK}).

    It is important to be aware that not all results
    that are valid for positive definite spaces are valid when the
    inner product is not positive definite.  For example, not all
    pseudo-hermitian operators are diagonalizable.  A good
    counterexample is precisely the operator $Q = \mathbf{J}^2$ above.
    In addition, not all eigenvalues are necessarily real.  In
    particular, a pseudo-hermitian operator may have complex
    eigenvalues that come in conjugate pairs.

    For our purposes, we will restrict consideration to
    pseudo-hermitian operators with real eigenvalues, of which $\mathbf{J}^2$
    will be the relevant example.  The domain
    of such an operator $A$ can always be decomposed into a direct sum
    of pairwise orthogonal subspaces, in each of which
    we can choose a basis such that $A$ has the
    so-called Jordan normal block form
    \begin{equation}
        \left(
            \begin{array}{cccc}
                \lambda & 1       &        &        \\
                        & \lambda & 1      &        \\
                        &         & \ddots & \ddots \\
                        &         &        & \lambda
            \end{array}
        \right) \label{JORDAN}
    \end{equation}
    and the inner product has the form
    \begin{equation}
        \pm\left(
            \begin{array}{cccc}
                 &        &        & 1       \\
                &         & 1      &        \\
                & \cdots         &        &        \\
                1        &         &        &
            \end{array}
        \right). \label{GRAM}
    \end{equation}
    Notice that only the first vector in the subspace is an
    eigenvector of $A$ with eigenvalue $\lambda$.
    If the Jordan block has dimension
    larger than $1$, this eigenvector is null.
    The vectors $v$ in this subspace are
    called principal vectors belonging to the eigenvalue $\lambda$ and
    satisfy
    \[
        (A - \lambda)^m v = 0
    \]
    for some integer $m \ge 1$.  The sequence of vectors $v_i$ spanning
    this subspace satisfying
    \[
        A v_{i} = \lambda v_i + v_{i-1}
    \]
    is called a Jordan chain.

    \section{BRST analysis of general product representations}

    We now generalize the method of section \ref{HALFHALF} to
    the analysis of a general product representation.

    In appendix \ref{J2ANALYSIS} we prove the result, important in
    what follows, that the Casimir operator $\mathbf{J}^2$ can be
    decomposed into Jordan blocks of dimension at most two.  Since
    the one and two-dimensional Jordan blocks of $\mathbf{J}^2$ are of the
    forms $\left(\lambda_i\right)$ and
    \begin{equation}
        \left(
            \begin{array}{cc}
                \lambda_i & 1 \\
                0 & \lambda_i
            \end{array}
        \right), \label{J2BLOCK}
    \end{equation}
    we can build a BRST operator in the
    product space of two arbitrary representations in terms of
    $\mathbf{J}^2$ as the orthogonal direct sum
    \begin{equation}
        Q = Q_{\lambda_1} \oplus Q_{\lambda_2} \oplus \cdots,
        \label{Q}
    \end{equation}
    where $Q_{\lambda_i}$ is defined on the principal vector subspace
    $V_{\lambda_i}$ belonging to the eigenvalue $\lambda_i$ of $\mathbf{J}^2$ by
    \begin{equation}
            Q_{\lambda_i} = \mathbf{J}^2|_{V_{\lambda_i}} -
            \lambda_i.
        \label{QI}
    \end{equation}
    Since $Q_{\lambda_i}$ consists of one- and two-dimensional blocks of the
    forms
    $\left(0\right)$ and
    \begin{equation}
        \left(
            \begin{array}{cc}
                0 & 1 \\
                0 & 0
            \end{array}
        \right), \label{QBLOCK}
    \end{equation}
    $Q$ is nilpotent and is a valid BRST operator.

    Taking the cohomology of $Q$ discards any Jordan
    blocks of dimension two in the decomposition of
    $\mathbf{J}^2$, so that all principal vectors of
    the reduced $[\mathbf{J}^2]$ will be eigenvectors.
    Equivalently, $[\mathbf{J}^2]$ can be decomposed into Jordan
    blocks of dimension $1$.

    Let us see how $Q$ affects the analysis of a
    general product representation.  As in the example of section
    \ref{HALFHALF}, positive spin $j$ representations may be
    present in the product in the form
    \begin{equation}
        \phi_{-j}^{(j)}
        \Jpmarrow \phi_{-j+1}^{(j)}
        \Jpmarrow \cdots
        \Jpmarrow \phi_{j}^{(j)}
        \Jpmarrow
           \boxed{
            \begin{array}{c}
                \phi_{j+1}^{(j)} \\
                \tilde\phi_{j+1}^{(j)}
            \end{array}
           }
        \Jpmarrow
            \boxed{
            \begin{array}{c}
                \phi_{j+2}^{(j)} \\
                \tilde\phi_{j+2}^{(j)}
            \end{array}
           }
        \Jpmarrow
        \cdots
        \label{BADREP}
    \end{equation}
    The boxes represent two-dimensional subspaces spanned by null
    vectors $\phi_{j+n}^{(j)}$ and $\tilde\phi_{j+n}^{(j)}$ on
    which
    $\mathbf{J}^2$ has the Jordan normal form (\ref{J2BLOCK})
    and $Q$ has the form (\ref{QBLOCK}).  We see that just as in
    the example of section \ref{HALFHALF},
    the ladder of states generated from the lowest weight $\phi_{-j}^{(j)}$
    consists of null states for $m > j$.

    Because of the additional states to the right of
    $\phi_{-j}^{(j)}$, this representation is not in the
    original class of irreducible
    positive or negative spin representations that we started
    out with.  In other words, the the original class of
    irreducible representations is not closed under
    multiplication, unless we can get rid of the extra states.

    Taking the cohomology with respect to $Q$ discards
    the boxed subspaces, and on the cohomology
    classes the representation has the familiar form
    \[
        [\phi_{-j}^{(j)}]
        \Jpmarrow [\phi_{-j+1}^{(j)}]
        \Jpmarrow \cdots
        \Jpmarrow [\phi_{j}^{(j)}].
    \]
    Henceforth, when we analyze product representations, it will
    always be assumed that we are working in the cohomology with respect to the
    associated operator $Q$.  This amounts to a redefinition of
    the product as the tensor product followed by the BRST
    reduction.

    \section{Generalized characters}

    The analysis of the decomposition of product representations
    may be greatly simplified by reformulating it as an algebraic
    problem in terms of characters.

    We would like the expression for the character of a representation
    to be invariant under the above
    BRST reduction to the cohomology of $Q$.
    Since the discarded Jordan subspaces have zero metric
    signature, an expression for the character of a
    group element $U$ that ignores these blocks is
    \begin{equation}
        \chi(U) = \sum_n \sig (V_{\lambda_n}) \lambda_n
        \label{CHARACTER}
    \end{equation}
    where $V_{\lambda_n}$ is the principal vector subspace
    corresponding to the eigenvalue $\lambda_n$ of $U$ and $\sig(\cdot)$
    denotes the signature.  Since the signature
    of a subspace is invariant under unitary transformations $V$ \cite{MALCEV}, this
    gives a basis invariant expression
    invariant under conjugation $U \to VUV^{-1}$.

    Using the following properties of the signature
    \begin{eqnarray}
        \sig (V \otimes W) &=& \sig V \cdot \sig W, \\
        \sig (V \oplus W) &=& \sig V + \sig W,
    \end{eqnarray}
    it follows that the characters satisfy the following important
    algebraic properties
    \begin{eqnarray}
        \chi (U_1 \otimes U_2) &=& \chi(U_1) \cdot\chi(U_2), \\
        \chi (U_1 \oplus U_2) &=& \chi(U_1) + \chi(U_2).
    \end{eqnarray}
    These are exactly the properties that make the characters
    useful for analyzing the decomposition of a product
    representation into a direct sum of irreducible representations.
    The first property ensures that the character of a product
    representation is simply the product of the characters of the
    individual representations.  In other words,
    \[
        \chi^{R_1\otimes R_2} = \chi^{R_1} \chi^{R_2}.
    \]
    The second property then ensures that if $R_1\otimes R_2 = \sum_i
    n_i R_i$, then
    \[
        \chi^{R_1\otimes R_2} = \sum_i n_i \chi^{R_i},
    \]
    where the weight $n_i$ denotes the degeneracy of the representation $R_i$
    in the product.  In the indefinite metric case, the weight $n_i$ will
    be negative if the inner product on $R_i$ is of opposite sign to the
    usual conventions as in (\ref{INNERPROD}), or equivalently, if the
    lowest weight state has negative norm squared.
    This is due to the inclusion of the signature in the
    definition (\ref{CHARACTER}) above.

    Specializing to $U = e^{i\theta J_z}$, we can now compute the
    characters of the negative dimensional representations.  We
    find, for $j < 0$
    \begin{eqnarray}
        \chi^{(j)}(\theta) &=&  e^{ij\theta} \,
           (1 - e^{i\theta} + e^{2i\theta} - \cdots) \nonumber\\
           &=& {e^{ij\theta}\over 1 + e^{i\theta}}.
            \label{CHARNEGATIVE}
    \end{eqnarray}
    Note that the singularity at $\theta = \pi$ is to be expected.  Indeed,
    as we remarked in section \ref{FINITE}, certain group elements in the conjugacy
    class of rotations by $\pi$ are not representable on the negative spin
    state space.  The singularity in
    the character reflects this fact.

    \section{General product decompositions}

    Let us first consider the combination of two negative spin
    representations.  For $j_1, j_2 < 0$, we have
    \begin{eqnarray}
        \chi^{(j_1)} \chi^{(j_2)} &=& e^{i(j_1 + j_2)\theta}
            \left({1\over 1 + e^{i\theta}}\right)^2\nonumber\\
            &=& e^{i(j_1 + j_2)\theta}
            \, {1\over 1 + e^{i\theta}} \,(1 - e^{i\theta} + e^{2i\theta} - \cdots)
               \nonumber\\
            &=& \chi^{(j_1 + j_2)} - \chi^{(j_1 + j_2 - 1)}
                    + \chi^{(j_1 + j_2 - 2)} - \cdots.
    \end{eqnarray}
    This then implies that, for $j_1, j_2 < 0$, we have
    \begin{equation}
        R^{(j_1)}\otimes R^{(j_2)} = \sum_{J = -\infty}^{j_1 + j_2}
        (-)^{j_1 + j_2 - J} R^{(J)}.
    \end{equation}
    The analysis of the product of positive and negative
    representations is most easily done by expanding in
    powers of $e^{i\theta}$ as in (\ref{CHARPOSITIVE}) and
    (\ref{CHARNEGATIVE}), performing the multiplication and
    collecting terms.  This is straightforward, and we state the
    results.

    When $j_1 > 0$, $j_2 < 0$ and $|j_2| > j_1$, we have
    \begin{equation}
        \chi^{(j_1)}\chi^{(j_2)} = \chi^{(-j_1 + j_2)} + \chi^{(-j_1 + j_2 + 1)}
                + \cdots + \chi^{(j_1 + j_2)},
    \end{equation}
    so that
    \begin{equation}
        R^{(j_1)}\otimes R^{(j_2)} = \sum_{J = -j_1 + j_2}^{j_1 + j_2}
            R^{(J)}.
    \end{equation}
    When $j_1 > 0$, $j_2 < 0$ and $|j_2| \le j_1$, and $j_1 - j_2$
    is integral, then
    \begin{eqnarray}
        \chi^{(j_1)}\chi^{(j_2)} &=& \chi^{(j_1 + j_2)} - \chi^{(j_1 + j_2 - 1)}
          + \cdots \pm \chi^{(0)} \nonumber\\
          & & {} + \chi^{-j_1 + j_2} + \chi^{-j_1 + j_2 + 1} + \cdots
             +\chi^{-j_1 - j_2 - 2},
    \end{eqnarray}
    so that
    \begin{equation}
        R^{(j_1)}\otimes R^{(j_2)} =
            \sum_{J=-j_1 + j_2}^{-j_1 - j_2 -2} R^{(J)}
            + \sum_{J = 0}^{j_1 + j_2}
           (-)^{j_1 + j_2 - J} R^{(J)}.
    \end{equation}
    When $j_1 > 0$, $j_2 < 0$ and $|j_2| < j_1$, and $j_1 - j_2$
    is half-integral, then
    \begin{eqnarray}
        \chi^{(j_1)}\chi^{(j_2)} &=& \chi^{(j_1 + j_2)} - \chi^{(j_1 + j_2 - 1)}
          + \cdots \pm \chi^{(-\half)} \nonumber\\
          & & {} + \chi^{-j_1 + j_2} + \chi^{-j_1 + j_2 + 1} + \cdots
              + \chi^{-j_1 - j_2 - 2},
    \end{eqnarray}
    so that
    \begin{equation}
        R^{(j_1)}\otimes R^{(j_2)} =
            \sum_{J=-j_1 + j_2}^{-j_1 - j_2 -2} R^{(J)}
            + \sum_{J = -\half}^{j_1 + j_2}
           (-)^{j_1 + j_2 - J} R^{(J)}.
           \label{HALFINTDIFF}
    \end{equation}
    These are the main results of this paper.

    \section{Dimensions}

    The state spaces of the negative spin representations are
    infinite-dimensional vector spaces.  However, a more useful
    concept of dimension may be obtained by defining the effective
    dimension of a representation $R$ to be the character
    $\chi^{(R)}(\mathbf 1)$ evaluated on the unit matrix.  While this
    coincides with the usual dimension in the case of positive
    spin representations, in the case of negative
    spin, taking $\theta = 0$ in (\ref{CHARNEGATIVE})
    gives a fractional dimension
    \[
        {e^0 \over 1 + e^0} = \half
    \]
    for all the negative spin representations.
    Since $\chi^{(R)}(\mathbf 1)$
    is, by our definition of the character, just the signature of the
    representation space, this may also be thought of as
    \[
        1 - 1 + 1 - 1 + \cdots = \half,
    \]
    which coincides with the Abel regularization $\lim_{z\to 1}\sum z^n$ of this
    alternating series.

    From the formulae of the previous section, we see that
    whenever a representation appears with inner product
    of sign opposite to the
    usual convention (\ref{INNERPROD}),
    we should think of its dimension as
    negative.

    Since the dimensions are defined in terms of the characters,
    the results of the previous sections guarantee that the
    dimensions behave in the expected way under multiplication and
    summation of representations.  For example, for $j_1, j_2 < 0$, we
    had
    \begin{equation}
        R^{(j_1)}\otimes R^{(j_2)} = \sum_{J = -\infty}^{j_1 + j_2}
        (-)^{j_1 + j_2 - J} R^{(J)}.
    \end{equation}
    The dimension of the tensor product on the left hand side is
    $\half\cdot\half = {1\over 4}$.  The sum of the dimensions on
    the right hand side is
    \[
        \half - \half + \half - \cdots = \half\, (1 - 1 + 1 - \cdots)
            = \half\cdot \half = {1\over 4},
    \]
    as expected.

    As another example, consider
    \[
        R^{({5\over 2})}\otimes R^{(-{3\over 2})}
          = R^{(1)} - R^{(0)} + R^{(-3)} + R^{(-4)}.
    \]
    The dimension of the tensor product on the left hand side is
    $6 \cdot \half = 3$.  Summing the dimensions of the representations
    on the right hand
    side, we obtain
    \[
        3 - 1 + \half + \half,
    \]
    again as expected.

    \section{Example: $R^{(1)}\otimes R^{(-\half)}$}

    We conclude by working out the following example from (\ref{HALFINTDIFF})
    \begin{equation}
        R^{(1)}\otimes R^{(-\half)} = R^{(\half)} - R^{(-\half)}.
        \label{EXAMPLER}
    \end{equation}
    Let us introduce the notation $\ket{m}_v \equiv e^{(1)}_m$ for the
    vector representation.
    From the above decomposition formula,
    we expect to find a spin $\half$ and a spin $-\half$
    lowest weight state in the product representation.  The spin
    $\half$ lowest weight state is just
    \[
        \ket{\hhalf, \textstyle{-\half}} \equiv \ket{-1}_v\otimes\ket{\hhalf}_g.
    \]
    This state is trivially annihilated by $J_-$.  Furthermore,
    using (\ref{JPLUSACTION}) we find
    \[
        J_+\ket{\hhalf, \textstyle{-\half}} = \sqrt 2\ket{0}_v
        \otimes\ket{\hhalf}_g + i \ket{-1}_v\otimes\ket{\textstyle{{3\over 2}}}_g,
    \]
    which is not null.  However, applying $J_+$ twice, we get the
    state
    \[
      (J_+)^2\ket{\hhalf, \textstyle{-\half}} = 2 \left(
        \ket{1}_v\otimes\ket{\hhalf}_g + i \sqrt{2}
        \ket{0}_v\otimes\ket{\textstyle{{3\over 2}}}_g -
        \ket{-1}_v\otimes\ket{\textstyle{{5\over 2}}}_g
      \right),
    \]
    which is null.

    A spin $-\half$ lowest weight state in the product is given by
    \[
        \ket{\textstyle{-\half}, \hhalf} \equiv i \sqrt 2 \ket{-1}_v
        \otimes\ket{\textstyle{{3\over 2}}}_g + \ket{0}_v\otimes
        \ket{\hhalf}_g.
    \]
    This lowest state has negative norm squared, as expected from
    the negative sign in (\ref{EXAMPLER}).  In addition, it is
    orthogonal to the spin $\half$ state $J_+\ket{\half, -\half}$
    with the same $m$-value.  By repeatedly applying $J_+$, we get
    an infinite tower of states filling out the spin $-\half$
    representation.

    Each of the subspaces with a specific value of $J_z$
    eigenvalue $m$ is three-dimensional.  One of these dimensions
    is taken up by the semi-infinite spin $-\half$ tower.  The
    remaining two dimensions belong to the spin $\half$
    representation, which has the form (\ref{BADREP}).  Taking the
    cohomology with respect to $Q$ gets rid of the unwanted
    higher $m$ states in the spin $\half$ representation, and we
    are left with the classes
    \[
        [\ket{\hhalf, \textstyle{-\half}}], \qquad [J_+\ket{\hhalf, \textstyle{-\half}}].
    \]
    Now $[J_+]^2 \,[\ket{\half, -\half}] = 0$, and we have an
    ordinary spin $\half$ representation on the cohomology.  The
    spin $-\half$ representation remains unaffected by taking the
    cohomology.

    The dimension of the tensor product is $3\cdot\half = {3\over
    2}$.  The sum of the dimensions of the decomposition on the
    right hand side is $2 - \half = {3\over 2}$, so the two sides
    match up as expected.

    \section{Conclusion}

    We showed that the representation theory of $su(2)$ may be
    extended to include a class of negative spin representations.
    We showed that arbitrary positive and negative representations
    may be multiplied and decomposed into irreducible
    representations after reducing the
    product to the cohomology of a certain nilpotent BRST-like operator
    $Q$.  We defined characters and effective dimensions for the negative spin
    representations and
    wrote down explicit formulae for the product decomposition
    in the general case.

    The program initiated here may be extended in various
    directions.  First, it would be interesting to work out
    the Clebsch-Gordan coefficients, 3-$j$ and 6-$j$ symbols in
    the general case.  Second, although we have argued that the
    negative spin Lie algebra representations do not exponentiate
    to representations of the whole group, it is possible to
    write down explicit forms for a subclass of finite $SU(2)$
    transformations (see \cite{MYSELF}).  It would be interesting
    to understand somewhat better the status of these
    transformations.

    Finally, it may be possible to apply some of the ideas developed
    here to a larger class of Lie algebras.

    \section*{Acknowledgments}

    The author would like to thank prof. Antal Jevicki and the Brown
    University Physics department for their support.

    \section*{Appendices}
    \appendix

    \section{BRST cohomology}
    \label{BRST}

    In the BRST formalism (see \cite{BRST}, \cite{HENNEAUX} and references therein),
    the analysis of physical states and
    operators is carried out in terms of an operator $Q$ that is hermitian
    and nilpotent.  In other words,
    \[
        Q^\dagger = Q, \qquad Q^2 = 0.
    \]
    States are called physical if they satisfy
    \[
        Q\ket{\phi} = 0,
    \]
    and are regarded as equivalent if they differ by a $Q$-exact
    state. In other words,
    \[
        \ket{\phi} \sim \ket{\phi} + Q \ket{\chi},
    \]
    where $\ket{\chi}$ is an arbitrary state.
    More formally, physical states are elements of the
    cohomology of $Q$, defined as the quotient vector space
    \[
        {\ker Q\, / \,\im Q}
    \]
    with elements
    \[
        [\ket{\phi}] \equiv \ket{\phi} + \im Q.
    \]
    The inner product on this quotient space may be defined in
    terms of the original inner product by
    noting that all elements of $\im Q$ are orthogonal to all
    elements of $\ker Q$, so that the induced inner product
    defined on equivalence classes in the cohomology by
    \[
        \braket {\phi + \im Q}{\phi' + \im Q} \equiv
        \braket{\phi}{\phi'}, \qquad \phi, \phi' \in \ker Q
    \]
    is well defined.

    A hermitian operator $A$ is regarded as physical if $[A, Q] =
    0$.  This ensures that $A$ leaves $\im Q$ invariant, so that
    the reduced operator $[A]$ defined on the cohomology classes by
    \begin{equation}
        [A]\,[\ket{\phi}] = [A\ket{\phi}]   \label{AREDUCED}
    \end{equation}
    is well-defined.

    \section{Jordan decomposition of $\mathbf{J}^2$}
    \label{J2ANALYSIS}

    In this appendix we prove that $\mathbf{J}^2$ can be decomposed
    into Jordan blocks of dimension at most two.

    First, note that, since $\mathbf{J}^2$ commutes with $J_z$,
    we can decompose $\mathbf{J}^2$ into Jordan blocks in each
    eigenspace of $J_z$.  Consider such a Jordan block on a subspace
    $V_m$ consisting of
    principal vectors of $\mathbf{J}^2$ belonging to an eigenvalue
    $j(j+1)$ and with
    $J_z$ eigenvalue $m$.  Taking any vector $v$ in $V_i$,
    by applying $J_-$ to it repeatedly we will eventually obtain
    zero, since our product representations are bounded below in
    $J_z$.  This gives a lowest weight state $J_-^k v$, and we can
    use (\ref{CASIMIR}) to obtain $j(j+1)$ in terms of the $J_z$ eigenvalue
    $m - k$ of this
    lowest weight state.  Since $m$ is integer or half-integer,
    the possible values of $j$ are also integer or
    half-integer, either positive or negative.  In the following,
    we shall take the positive solution $j>0$.

    Now consider the sequence of subspaces
    \[
        V_m \stackrel{J_-}{\longrightarrow} V_{m-1} \stackrel{J_-}{\longrightarrow}
        V_{m-2} \stackrel{J_-}{\longrightarrow} \cdots.
    \]
    Since $\mathbf{J}^2$ commutes with $J_-$, we see that
    $\mathbf{J}^2$ takes each of the subspaces $V_{i}$ to
    itself.  Furthermore, as long as $i \not\in \left\{-j, j + 1\right\}$, $J_-$
    cannot change the dimension of these subspaces since that
    would imply that $\ker J_-$ is not empty, so there would be
    lowest weight states at values of $i$ inconsistent with $j$.
    In other words, the dimensions of the above sequence of spaces
    $V_i$ can at most jump at $i \in \left\{-j, j + 1\right\}$.
    As a corollary, the above sequence terminates at either $i =
    -j$ or $i = j + 1$.

    Furthermore, $\mathbf{J}^2$ consists of a single
    Jordan block on each of the subspaces $V_i$.
    By assumption, this is true for the first element $V_m$
    of the sequence.  In general, assume that $\mathbf{J}^2$ consists of a
    single Jordan block on $V_i$ and consider $V_{i-1}$.  Since from (\ref{CASIMIR})
    we have that $J_+ V_{i-1} = J_+ J_- V_i = (\mathbf{J}^2 - J_z^2 +
    J_z) V_i$, we see that $J_+ V_{i-1} \subseteq V_i$.  Now if $\mathbf{J}^2$
    were to consist of more than one Jordan block on $V_{i-1}$, each
    of these blocks would contain an eigenvector of $\mathbf{J}^2$.
    Since $J_+$ commutes with $\mathbf{J}^2$, all these
    eigenvectors will be taken by $J_+$ to eigenvectors in $V_i$,
    of which there is only one by assumption.
    Therefore $J_+$ is not one to one, its
    kernel on $V_{i-1}$ is nonempty, and there is a highest weight
    state in $V_{i-1}$, which is inconsistent with $j$ unless
    $i \in \{-j, j+1\}$.  This proves the assertion when
    $i \not\in \{-j, j+1\}$.

    Now consider the case $i = j + 1$.  The case $i = -j$ is similar.
    Choose $\tilde v_{j}\in V_{j}$ such
    that $J_+ \tilde v_{j} = 0$.  Since $V_{j} = J_- V_{j+1}$,
    there is a $\tilde v_{j+1}\in V_{j+1}$ such that $\tilde v_{j} = J_- \tilde
    v_{j+1}$.  Then $0 = J_+ J_- \tilde v_{j+1} = (\mathbf{J}^2 - J_z^2 + J_z)
    \tilde v_{j+1}$, which implies that $\tilde v_{j+1}$ is an eigenvector of
    $\mathbf{J}^2$ and therefore is proportional to
    the unique eigenvector $v_{j+1}$ in $V_{j+1}$.
    Now note that $J_+ v_{j}$ cannot be zero for all eigenvectors $v_{j}$ in
    $V_{j}$, because if that were the case, then by the
    above argument there would be more than one linearly independent eigenvector of
    $\mathbf{J}^2$ in $V_{j}$.  Therefore we can find a $v_{j}$ such that $J_+ v_{j} =
    v_{j+1}$.
    Then $J_- J_+ v_{j} = (\mathbf{J}^2 - J_z^2 - J_z) v_{j}
    = \left[j(j+1) - j(j+1)\right] v_j = 0$,
    or $J_- v_{j+1} = 0$.  But we had $0\ne\tilde v_j = J_- \tilde
    v_{j+1}$, and $\tilde v_{j+1} \propto v_{j+1}$, which is a
    contradiction.  This proves the assertion when $i = j + 1$.

    We have proved that $\mathbf{J}^2$ consists of a single Jordan
    block on each $V_i$.  This means that each $V_i$ contains one and only one
    eigenvector.  Consequently, since elements in the kernel of
    $J_-$ are automatically eigenvectors, the dimension of the
    $V_i$ can be reduced by at most one at each of the two transition
    points
    $V_{j+1} \stackrel{J_-}{\longrightarrow} V_j$ and
    $V_{-j} \stackrel{J_-}{\longrightarrow} V_{-j-1}$.  Since the
    sequence terminates at either $V_{j+1}$ or $V_{-j}$, the
    initial space $V_m$ can be at most two-dimensional.  This
    completes the proof that the Jordan blocks of $\mathbf{J}^2$ are at most
    two-dimensional.


\begin{thebibliography}{99}
        \bibitem{ELLIOTT}
            J.P. Elliott and P.G. Dawber,
            \textit{Symmetry in Physics}, Volumes 1 and 2, Macmillan
            1979;
            J.F. Cornwell, \textit{Group Theory in Physics -- An
            Introduction}, Academic Press, London 1997.
        \bibitem{MYSELF}
            A. van Tonder, \textit{Ghosts as Negative Spinors}, hep-th/0207110.
        \bibitem{BOGNAR}
            J. Bognar, \textit{Indefinite Inner Product Spaces},
            Springer-Verlag, Berlin 1974.
        \bibitem{MALCEV}
            A.I. Mal'cev, \textit{Foundations of Linear Algebra},
            W.H. Freeman and Company 1963.
        \bibitem{JACOBZYK}
            L, Jac\'obczyk,
            \textit{Ann. Phys.} \textbf{161}, 314 (1985);
            G. Scharf, \textit{Finite Quantum Electrodynamics}, Second Edition
            pp. 149-156, Springer-Verlag 1995.
        \bibitem{BRST}
            I.V. Tyutin, \textit{Lebedev preprint} FIAN \textbf{39} (1975),
            unpublished;
            C. Becchi, A. Rouet and R. Stora, \textit{Ann. Phys.} \textbf{98} (1976),
            28.
        \bibitem{HENNEAUX}
            M. Henneaux, \textit {Phys. Rep.} \textbf{126} (1985), 1.
    \end{thebibliography}
\end{document}